\documentclass[aps,prl,twocolumn]{revtex4}

\usepackage{epsfig,pstricks}
\usepackage{psfrag}

\usepackage{amsmath,amssymb}
\usepackage{color}

\def\CC{{\rm\kern.24em \vrule width.04em height1.46ex depth-.07ex
\kern-.30em C}}
\def\P{{\rm I\kern-.25em P}}
\def\NN{{\rm I\kern-.25em N}}
\def\RR{{\rm
         \vrule width.04em height1.58ex depth-.0ex
         \kern-.04em R}}
\def\id{{\rm 1\kern-.22em l}}
\def\ZZ{{\sf Z\kern-.44em Z}}
\def\tr{{\rm tr}\;}
\newtheorem{pdef}{Definition}[section]

\newenvironment{eqblock}[2]{\beq\label{#2}\begin{array}{#1}}{\end{array}
                                \eeq}
\newenvironment{neqblock}[1]{\[\begin{array}{#1}}{\end{array}\]}

\newcommand{\ketbra}[1]{\ensuremath{| #1 \rangle \langle #1 |}}

\newcommand{\beqb}{\begin{eqblock}}
\newcommand{\eeqb}{\end{eqblock}} 
\newcommand{\nbeqb}{\begin{neqblock}}
\newcommand{\neeqb}{\end{neqblock}} 

\newcommand{\beq}{\begin{equation}}
\newcommand{\beqa}{\begin{eqnarray}}
\newcommand{\eeq}{\end{equation}}
\newcommand{\eeqa}{\end{eqnarray}}
\newcommand{\nbeqa}{\begin{eqnarray*}}
\newcommand{\neeqa}{\end{eqnarray*}}

\newcommand{\ket}[1]{| #1 \rangle}

\newcommand{\Matrix}[2]{\left( \begin{array}{#1} #2 \end{array}
  \right)}

\def\DJo{$\;$\kern-.4em \hbox{D\kern-.8em\raise.15ex\hbox{--}\kern.35em okovi\'c}}

\begin{document}

\title{Upper bound for SL-invariant entanglement measures of mixed states}
\author{Andreas Osterloh}
\affiliation{Institut f\"ur Theoretische Physik, 
         Universit\"at Duisburg-Essen, D-47048 Duisburg, Germany.}
\email{andreas.osterloh@uni-due.de}
\begin{abstract}
An algorithm is proposed that serves to handle full rank density matrices, when coming from a
lower rank method to compute the convex-roof. This is in order to calculate an upper bound for
any polynomial SL invariant multipartite entanglement measure E. Here, it is exemplifyed how this
algorithm works, based on a method for calculating convex-roofs of rank two density matrices.
It iteratively considers the decompositions of the density matrix into
two states each, exploiting the knowledge for the rank-two case. 
The algorithm is therefore quasi exact as far as the two rank case is concerned, and it also gives 
hints where it should include more states in the decomposition of the density matrix.
Focusing on the threetangle,
I show the results the algorithm gives for two states, one of which being the $GHZ$-Werner state, 
for which the exact convex roof is known. It overestimates the threetangle in the state, 
thereby giving insight into the optimal decomposition the $GHZ$-Werner state has.
As a proof of principle, I have run the algorithm for the threetangle on the 
transverse quantum Ising model.  
I give qualitative and quantitative arguments why the convex roof should be close to the upper
bound found here.
\end{abstract}

\maketitle

\section{Introduction}
The quantification of entanglement is a central issue in physics but it is the convex roof construction
which renders this subject so difficult in practice for arbitrary mixed states. This problem was shown 
to be solvable for the $SL$-invariant concurrence of two qubits\cite{Wootters98} 
and could then be extended to every entanglement measure which is a 
homogeneous polynomial of degree two in the coefficients of the wave function, 
as for the concurrence\cite{Uhlmann00}. For higher homogeneous degree this was not possible, since 
the entanglement measure can not be written any more as a simple expectation value of 
some antilinear operator\cite{OS04,OS05,DoOs08}.
It was pointed out, however, in Refs.~\cite{LOSU,KENNLINIE} that 
a numerically exact solution is available in principle for the zero-simplex: each polynomial 
$SL$-invariant measure $E_{orig}$ of homogeneous degree $D$ leads to an order $D$ polynomial 
equation in a single complex variable $z$ via 
$E_{orig}(\psi_1+z \psi_2)=0$. For each convex combination of these $D$ solutions the 
entanglement measure $E$ is hence zero. Therefore this simplex made of $D$ points has been termed the
zero-simplex.  
Also for higher ranks such polynomials do exist in however more than a single variable\cite{KENNLINIE},
leading to a manifold of such zero-simplices, whose convexification gives rise to
a zero-manifold.
Exact convex roofs have been published for mixed states of rank two for three qubits\cite{LOSU,EOSU} and the threetangle $\tau_3$\cite{Coffman00}, but also for
higher ranks of special three qubit density matrices\cite{Jung09,HigherRankTau3}.
Further exact results for three qubits have been obtained when imposing certain conditions on the states in 
Ref.~\cite{Eltschka2012,Siewert2012}. In these works, the symmetry that is respected by 
three qubit GHZ states was imposed
to reduce the parameters in the density matrix to two, and an exact convex roof expression has been obtained. 
For arbitrary states this constitutes a lower bound to $\sqrt{\tau_3}$\cite{Eltschka2014},
making this approach similar to a witness approach (see Ref.~\cite{EltschkaS12,EltschkaS13} and refs. therein). 
Rather recently, an algorithm that finds an upper bound for SL-invariant entanglement measures has been published. 
The main idea behind this procedure is, to find pure states in the zero-simplex and obtain another pure state 
as prolongation of the straight line that interconnects the actual density matrix $\rho$ 
with the equal mixture of some extremal states out of the zero-simplex\cite{Rodriquez14}.
The algorithm has certain advantages, in particular it is basically taking into account the 
full range of $\rho$; but it lacks 
the knowledge from the rank two setting of Ref~\cite{LOSU,KENNLINIE}
and is a purely stochastic approach.

Here, inspired by the method used in Ref.~\cite{Jungnitsch11}, 
I propose an algorithm based on precisely this. 
I will restrict myself to the superpositions of two states each
and give an upper bound of its entanglement content iteratively. 
To this end, I will focus only on functions $E$ of effective homogeneous degree two, 
this means $E=E^{2/D}_{orig}$. Below I will analyse the three-tangle $\tau_3$,
which is a homogeneous polynomial of degree $D=4$, and hence $E=\sqrt{\tau_3}$.
but the algorithm also gives an upper bound if this restriction is relaxed.
The disadvantage occurs whenever more than two states are to be combined in a convex roof.

This work is laid out as follows. In the next section I describe the algorithm in detail. Then,
I demonstrate that the restriction included in the algorithm is relevant of course, but
can be removed by hand in the specific cases I consider. At the end, I run the algorithm on the 
threetangle of the transverse quantum Ising model as a proof of principles and draw my conclusions 
and give an outlook on possible future directions of research.

\section{Description of the algorithm}\label{Algo}

Suppose we have a density matrix of rank $r$ 
\beq
\rho=\sum_{j=1}^r \lambda_j\ketbra{\psi_j}
\eeq
with $\lambda_1\leq\lambda_2\leq ... \leq \lambda_r$, and $\ket{\psi_j}$ are the corresponding orthonormal 
eigenstates of the density matrix.
The first step is to take only into consideration a rank-two part of the density matrix, e.g.
\beq
\rho_{1,2}^{(1)}(p_{\rm i})=p_{\rm i}\ketbra{\psi_{r-1}}+(1-p_{\rm i})\ketbra{\psi_r}
\eeq
with $p_{\rm i}=\lambda_{r-1}/(\lambda_{r-1}+\lambda_r)$.
Next we search for all solutions of the zero-simplex, given by the equation
\beq
E_{orig}[\sqrt{p_{\rm i}}\ket{\psi_{\rm i}}+\sqrt{1-p_{\rm i}}e^{i \varphi}\ket{\psi_{\rm f}}]=0
\eeq
for $\ket{\psi_{\rm i}}=\ket{\psi_{r-1}}$ and $\ket{\psi_{\rm f}}=\ket{\psi_r}$ and 
corresponding eigenvalues $\lambda_{\rm i}$ and $\lambda_{\rm f}$,
which has $D$ solutions for any $SL$-invariant measure of entanglement $E$ which 
is originally a polynomial of homogeneous degree $D$.
Hence, we look for solutions to the equation in $z$
\beq
0=\left\{
\begin{array}{c}
E_{orig}[\ket{\psi_{\rm i}}+z \ket{\psi_{\rm f}}]\\
E_{orig}[\ket{\psi_{\rm f}}+z \ket{\psi_{\rm i}}]
\end{array}\label{zero-simplex}
\right.
\eeq
and make sure that we have the complete set of $D$ solutions, which we call $z_{0;j}$ for $j=1,\dots , D$.
The corresponding state to the zero $z_{0;j}$ of $E$ is termed $\ket{\Psi_{z_{0;j}}}$. 
Every such zero corresponds to the probabilities $p_{0;j}=1/(1+\left|z_{0,j}\right|^2)$ and 
$\tilde{p}_{0;j}=1-p_{0;j}=\left|z_{0,j}\right|^2/(1+\left|z_{0,j}\right|^2)$, which sum up to $1$.
Thus, the state is
\beq\label{correspondence}
\ket{\Psi_{z_{0;j}}}=\sqrt{p_{0;j}}\left[\ket{\psi_{\rm i}}+z_{0;j} \ket{\psi_{\rm f}}\right]\; .
\eeq
It must be mentioned that 
for states with $E[\ket{\psi}]=0$ the set of solutions of $E_{orig}[\ket{\phi}+z \ket{\psi}]$ 
will be reduced by at least one. Hence one either has to look for solutions to the second choice 
in~\eqref{zero-simplex}
or one has to set the missing solutions to $z_0=\infty$.
\begin{figure}
\centering
\psfrag{psi}[b]{$\vec{r}\hat{=}\psi(\vec{r})$}
\psfrag{2p-1}[b]{$2p-1$}
\psfrag{q}[b]{$\vartheta$}
\psfrag{f}[b]{$\varphi$}
  \includegraphics[width=.9\linewidth]{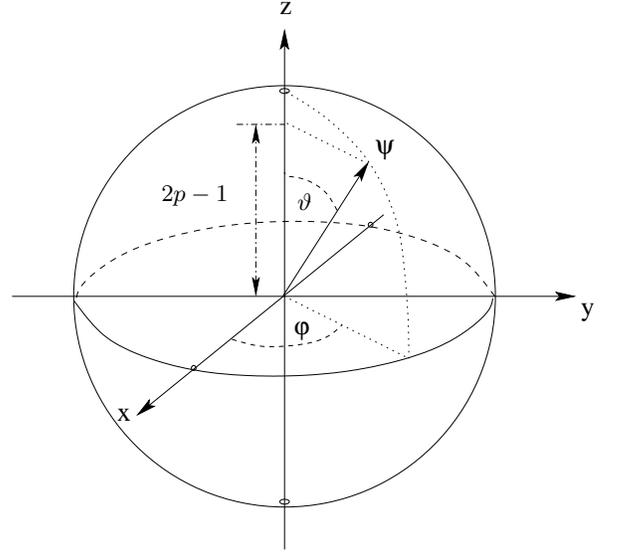}
  \caption{The Bloch sphere with radius $|\vec{r}|=1$ and the standard parametrization of its surface
in the angles $\vartheta=\theta$ and $\varphi$ of the vector $\vec{r}$. This vector corresponds to
the pure state $\ket{\psi(\vec{r})}=\cos{\frac{\vartheta(p)}{2}}\ket{\psi_{\rm i}}+ \sin{\frac{\vartheta(p)}{2}}e^{i\varphi} \ket{\psi_{\rm f}}=\sqrt{p}\ket{\psi_{\rm i}}+ \sqrt{1-p}e^{i \varphi}\ket{\psi_{\rm f}}$. The $r_z$ component is related to the probability $p$ via $r_z=2p-1$.}
  \label{blochsphere}
\end{figure}
To proceed further, it is instructive to use the Bloch-sphere (see fig.~\ref{blochsphere}), and the state
\beq
\ket{\psi(\vec{r})}=\cos{\frac{\vartheta(p)}{2}}\ket{\psi_{\rm i}}+ \sin{\frac{\vartheta(p)}{2}}e^{i\varphi} \ket{\psi_{\rm f}}\; ,
\eeq
with $\cos(\vartheta(p))=2p-1$,
that corresponds to the vector
\beq
\ket{\psi(\vec{r})}\hat{=}\vec{r}=\Matrix{c}{2\sqrt{(1-p)p}\,\cos{\varphi}\\
                                      2\sqrt{(1-p)p}\,\sin{\varphi}\\  
                                        2p-1}\; .
\eeq
Observe further that the solutions to the zero-simplex, following equation \eqref{correspondence}, become
\beqa
z_{0;j}&=&\sqrt{\frac{1-p_{0;j}}{p_{0;j}}}e^{i\varphi}\\
\Leftrightarrow
Z_{0;j}&:=&p_{0;j}z_{0;j} = \sqrt{p_{0;j}(1-p_{0;j})}e^{i\varphi}
\eeqa
Hence, the zero-simplex has an overlap with the central axis of the bloch sphere {\em iff}
\beq
\sum_{j\in J} \mu_j Z_{0;j}=0
\eeq
for convex combinations with weights $\mu_j$, where the intersection point is then given by
\beq
\sum_j \mu_j p_{0;j}=p\; .
\eeq
Here, $J$ is an index representing the $D$ elements of the zero simplex.
Various faces of the zero-simplex may lead to different intersection points, such that 
one obtains at the end an interval $[p_{min},p_{max}]$ of intersections with the central axis of
the Bloch-sphere.  
This has to be compared with the value for $p_{\rm i}$. 
If $p_{\rm i}$ is outside the interval $[p_{min},p_{max}]$, then either the state $\ket{\psi_{\rm i}}$ for $p_{\rm i}>p_{max}$ 
with weight $\widetilde{p}_{\rm i}=(p_{\rm i}-p_{max})/(1-p_{max})$, or
the state $\ket{\psi_{\rm f}}$ for $p_{\rm i}<p_{min}$ and weight 
$\widetilde{p}_{\rm f}=(p_{min}-p_{\rm i})/p_{min}$
will have survived in the rank-two density matrix.
The weights have to be chosen such that $E[\rho_{1,2}(p_i)]=E[\widetilde{p}_{\rm i/f}\ket{\psi_{\rm i/f}}]=
\widetilde{p}_{\rm i/f}E[\ket{\psi_{\rm i/f}}]$ for an entanglement measure of homogeneous degree $2$.
It is assumed that $E$ will have a linear interpolation between $[0,p_{min}]$
and $[p_{max},1]$ to give the result
\beq
E[\rho_{1,2}(p_i)]=E[\ket{\psi_s}]\left\{
\begin{array}{ccc}
(1-\frac{p_i}{p_{min}}) && p_i<p_{min}\\
&\mbox{for}&\\
\frac{p_i-p_{max}}{1-p_{max}} && p_i>p_{max}
\end{array}\right.
\eeq
If $E$ will be partially strictly convex on the interval - 
which however seldomly occurs
if the measure has a homogeneous degree of at most $2$ (see however Ref.~\cite{OsterlohNineWays15}) - 
this linear interpolation will lead to an upper bound of $E[\rho_{1,2}(p)]$ already for rank two;
otherwise it describes the convex roof $\hat{E}$ of the state $\rho_{1,2}(p)$ exactly.

This translates into
\beq
\widetilde{\lambda}_{\rm f}=\lambda_{\rm f}-\lambda_{\rm i}\frac{1-p_{min}}{p_{min}}\; ;
\eeq
for $0<p_{\rm i}<p_{min}$ and 
\beq
\widetilde{\lambda}_{\rm i}=\lambda_{\rm i}-\lambda_{\rm f}\frac{p_{max}}{1-p_{max}}\; .
\eeq
for $p_{max}<p_{\rm i}<1$ as weight inside the density matrix $\rho$.
I use $\widetilde{\lambda}_s:=\widetilde{\lambda}_{\rm i/f}$ 
and $\ket{\psi_s}:=\ket{\psi_{\rm i/f}}$ for the surviving species.

The remaining density matrix can then be written as
\beq
\rho^{(1)}=\frac{1}{\widetilde{\lambda}_s+\sum_{j=1}^{r-2} \lambda_j}
\left[\widetilde{\lambda}_s\ketbra{\psi_s}+\sum_{j=1}^{r-2} \lambda_j\ketbra{\psi_j}\right] 
\eeq
and we obtain the factor $f_1=\widetilde{\lambda}_s+\sum_{j=1}^{r-2} \lambda_j$ 
for the calculation of the final value of $E$ at the end.

It must be stressed that 
if the value of $p$ falls inside the zero-simplex, $p_{min}<p_i<p_{max}$,
both states would be taken out of the mixture, and therefor
\beq
\rho^{(1)}=\frac{1}{\sum_{i=1}^{r-2} \lambda_j}\sum_{j=1}^{r-2} \lambda_j\ketbra{\psi_j}\; .
\eeq
If this happens, this may be a hint that more than two states have to be used
inside the decomposition. 
Then, additional eigenstates could be added 
until the extremal points $p_{min}$ or $p_{max}$ would be reached. This idea however is not included in the algorithm
so far and could be one topic of future investigations.

If there does not exist an intersection point of the centerline with the zero-simplex,
choose the state $\psi_{off}$ such that its connecting line with $\rho$
hits the zero-simplex and that minimizes $\widetilde{\lambda}_s E[\psi_{\rm off}]$;
one can reduce the minimization to $\widetilde{\lambda}_s$, having in mind that both minima might not coincide,
hence giving an upper bound already here. Since $\psi_{\rm off}$ contains two eigenstates of $\rho$,
this leads effectively to having three eigenstates in the decomposition. 

We successively proceed with this algorithm, unless two final states are left in the mixture.
We then finally have
\beq
E[\rho]\leq E[\prod_{k=1}^{n-1} f_k \rho^{(n)}_{1,2}(p)]=
\prod_{k=1}^{n-1} \left(f_k^{D/2}\right)E[\rho^{(n)}_{1,2}(p)]
\eeq
if the homogeneous degree of $E$ is $D$. The so obtained value has to be joined to the upper bound of $E[\rho]$.
At the end one has to take the minimum of this set of values.

\section{Benchmarks} \label{bench}

\begin{figure}
\centering
  \includegraphics[width=.9\linewidth]{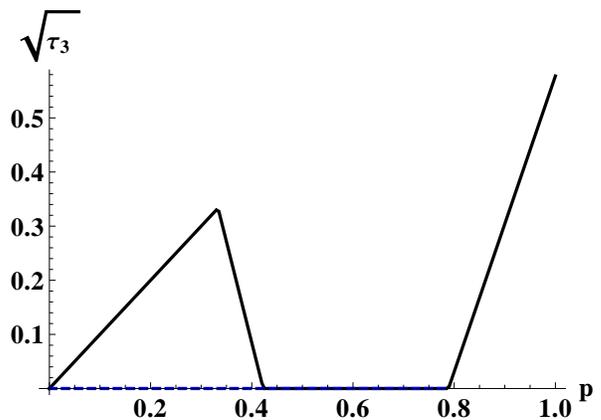}
  \caption{The figure produced by the algorithm (black curve) is shown together with the convexified curve (blue dashed curve),
which gives the correct upper bound to the convex roof. It hence is zero up to the value of $p\approx 0.788675$ and begins to linearly increase to
finally reach $\sqrt{\tau_3}[\ket{\varphi}]=1/\sqrt{3}$. At the value of $p=3/4$ it is hence zero, confirming the result of 
Ref.~\cite{Eltschka2014}. The minimal value of $p\approx 0.788675$ is a result of this work.}
  \label{fig:W-type}
\end{figure}

In order to demonstrate how well the algorithm works, we take two benchmark-states, 
where we know the entanglement measure exactly, and focus on the 
threetangle $\sqrt{\tau}_3$\cite{Coffman00,ViehmannII}.
Since the algorithm is based on the steps one has to do for a rank-two density matrix, 
the algorithm gives precisely the convex roof, whenever an affine exact solution is known 
for rank-two states.
Therefore, the first benchmark states are chosen to be the non-trivial $W$ states from Ref.~\cite{Eltschka2014}
\beqa
\rho_{W-like}&=&p\; \pi_{\ket{\varphi}}+\frac{1-p}{2}(\pi_{\ket{000}}+\pi_{\ket{111}})\label{1st:W}\\
&=&p\;\pi_{\ket{\varphi}}+\frac{1-p}{2}(\pi_{\ket{GHZ_+}}+\pi_{\ket{GHZ_-}})\label{2nd:W}
\eeqa
with the states
\beqa
\ket{\varphi}&=&\frac{1}{\sqrt{6}}(\ket{001}+\ket{010}+\ket{011}\nonumber\\ 
&&\quad +\ket{100}+\ket{101}+\ket{110})\\
\ket{GHZ_{\pm}}&=&\frac{1}{\sqrt{2}}(\ket{000}\pm \ket{111})
\eeqa
and the definition $\pi_{\ket{\psi}}=\ketbra{\psi}$. 
For these states a decomposition into three eigenstates is necessary 
for the construction of the convex roof.
Therefore, the bare algorithm leads to the straight line connecting $\sqrt{\tau_3}[\ket{\varphi}]=1/\sqrt{3}$ 
with zero.
We can however manipulate the eigenbasis a bit to obtain e.g. the case in \eqref{2nd:W}.
The graph produced by the algorithm for this eigenbasis is shown in fig.~\ref{fig:W-type}.
The square-root of the threetangle of this state vanishes at least until $p\approx 0.788675$. 
The upper bound then consequently linearly increases up to the value of $\sqrt{\tau_3}[\ket{\varphi}]=1/\sqrt{3}$. 
At the value of $p=3/4<0.788675$ it is hence zero, 
confirming this result of Ref.~\cite{Eltschka2014}.

\begin{figure}
\centering
  \includegraphics[width=.9\linewidth]{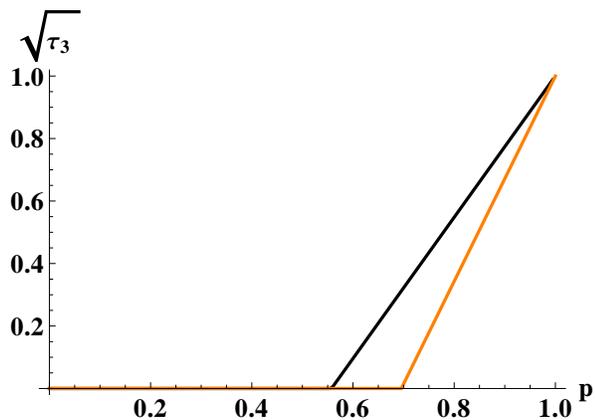}
  \caption{The figure produced by the algorithm (upper black curve) for the modified basis is shown 
together with the convex roof of $\sqrt{\tau_3}$ (lower orange curve)\cite{Siewert2012}.}
  \label{fig:GHZ-Werner}
\end{figure}

The next benchmark states are the generalized Werner states\cite{Werner89,Murao98} 
including the three-qubit $GHZ$ state 
\beq\label{GHZ-Werner}
\rho_{GHZ,{\rm Werner}}=p\,\pi_{\ket{GHZ}} + \frac{1-p}{8} \id\; .
\eeq
Also here, an application of the algorithm to an obvious eigenbasis of $\rho_{GHZ,{\rm Werner}}$ gives a simple straight 
line connecting $\sqrt{\tau_3}[\ket{GHZ}]=1$ at $p=1$ with $0$ at $p=0$. 
However, we can also here modify the eigenbasis
in that it contains $W$-states whose coefficients are suitable zeros of $z^3=1$. The result is shown in fig.~\ref{fig:GHZ-Werner}.
It definitely overestimates the threetangle in the state, 
but offers an upper bound that vanishes until $p_0\approx 0.55750$,
whereas the exact value is $p_0\approx 0.69554$.
The correct decomposition vectors of $\rho_{GHZ,{\rm Werner}}$ are made out of 
all of its eigenstates\cite{Siewert2012}. They could be seen in an extension of this algorithm 
to include up to three states in the decomposition.

It is clear that a slight (e.g. experimental) disorder can be dealt with: one has to take the minimum 
of the eigenvalues in the admixture where we wish to have equal eigenvalues. Then one has
\beq
\rho=p\,\pi_\psi +\Omega_{min} (1- p)\rho_0 +2\,\delta\!\rho_0\approx p\,\pi_\psi+(1-p)\rho_0
\eeq
with $\Omega_{min}\lesssim 1$, $2\,\tr \delta\!\rho_0= (1-\Omega_{min})(1-p)$, and 
\beq
\rho_1:=\frac{p}{p+\Omega_{min}(1-p)}\pi_{\psi} + \frac{\Omega_{min}(1-p)}{p+\Omega_{min}(1-p)}\rho_0\; .
\eeq
If the eigenstates in the degenarate $\rho_0$ do not carry entanglement, then one is left with 
$E[\rho]\leq (p+\Omega_{min}(1-p))E[\rho_1]$.

\section{Upper bound for the Ising model}\label{ising}
  
\begin{figure}
\centering
  \includegraphics[width=.9\linewidth]{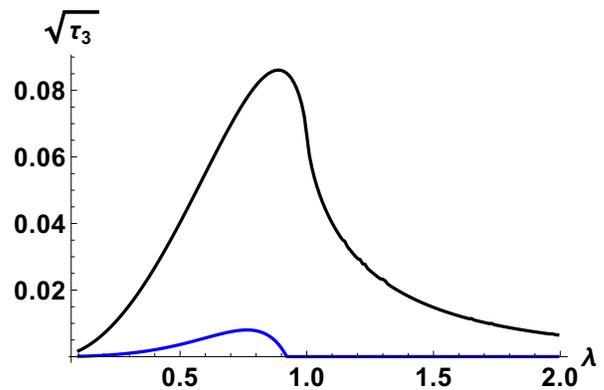}
  \caption{The upper bound coming out of the proposed algorithm is shown for the transverse quantum Ising model 
as a function of $\lambda$. 
I have furthermore calculated the lower bound of Ref.~\cite{Eltschka2014}. 
It is non-zero only for $0\leq\lambda\lesssim0.9206$ and stays considerably below the upper bound. 
Form $\lambda>0.9206$ it is zero. 
In particular, it is also the trivial lower bound 
in the vicinity of the critical point $\lambda=1$.}
  \label{fig:Tau3-Ising-111}
\end{figure}

In order to demonstrate the algorithm at work, we face towards an upper bound for the threetangle in the Ising model
in a transverse field
\beq\label{Ising}
H_{\rm Ising}= \sum_i (2\lambda S_i^x S_{i+1}^x + S_i^z)\; ,
\eeq
where the spin matrices $S_i^a=\sigma_i^a/2$ for $a=x,z$ enter, and $\lambda$ is a dimension-less 
interaction parameter; $\sigma^a$ are the Pauli matrices. 
Periodic boundary conditions are imposed and the model is considered 
in the thermodynamic limit.
It has a quantum phase transition at $\lambda=1$.
For the ground state of this integrable model\cite{PFEUTY,LIEB}, 
I use the expressions for the expectation values~\cite{Hofmann14} to calculate 
the upper bound to the nearest neighbor (adjacent sites) threetangle $\sqrt{\tau_3}$ 
given by this algorithm.
The result can be seen in fig.~\ref{fig:Tau3-Ising-111}.
The curve has, except the smaller range, an astonishing similarity to the 
nearest neighbor concurrence\cite{OstNat,Osborne02}: it has a maximum which is situated at about $\lambda=0.9$
and decays to zero for $\lambda\to\infty$.
With the genuine three party negativity\cite{Hofmann14}, which measures besides the $GHZ$ entanglement the
non-biseparable $W$ entanglement,  
also the magnitude is in the same range of $\sqrt{\tau_3}\sim 0.09$.
I also calculated the lower bound of Ref.~\cite{Eltschka2014}, which is positive but 
stays cosiderably below the upper bound for $0\leq\lambda\lesssim0.9206$. For
$\lambda>0.9206$ it is zero. In particular, it is the trivial lower bound 
in the vicinity of the critical point $\lambda=1$. So there is no hint towards
a finite three-tangle from this lower bound around the quantum critical point. 
It is highly probable that 
the eigenstates of the Ising model are too far away from the GHZ symmetric point
because I can almost infer from this upper bound a non-zero threetangle because 
the sum of the six smallest eigenvalues is below $0.029$ for all $\lambda$, 
and hence much smaller than the detected threetangle. The state with the third-largest weight
is biseparable and does not give a considerable decrease regarding the threetangle; 
considering the sum of the remaining eigenvalues, one ends up with a value which is smaller 
than $0.0056$ for all $\lambda$.
It is furthermore astonishing how well 
this upper bound fits into the 
genuine three-partite negativity of neighboring sites\cite{Hofmann14}. 
Further studies into this direction 
must clarify these points.

\section{Conclusions}\label{concl}
I have presented an algorithm that calculates an upper bound of $SL$-invariant entanglement measures $E$,
exploiting the knowledge for the rank-two case of Refs.~\cite{LOSU,KENNLINIE}. 
It is straightforward to include in the algorithm decompositions of three or more eigenstates
as soon as they are available.
It is thereby a method 
which is easily extendable to more than two states in the decomposition.
In part, however, the algorithm 
already contains three states in the decomposition of the density matrix and suggestions where 
more states will be necessary.
This opens up a method for experimentally testing the content of an arbitrary SL-invariant entanglement measure,
as the threetangle, in the given states.

I demonstrate the algorithm on various benchmark states for three qubits, 
where the exact convex roof is known~\cite{Siewert2012,Eltschka2014}. 
To this end, a modification of the eigenbasis of the states is shown to be obligatory in order to give 
a reasonable result.
This modification is shown to be stable as far as e.g. small experimental errors are concerned.
Whereas for the ``nontrivial W-state'' from Ref.~\cite{Eltschka2014} it already is sufficient to express the
product basis by a basis of maximally entangled $GHZ$-states, no such form exists for 
the $GHZ$-Werner state. 
Here, rewriting the product basis in a basis of $W$-states is not sufficient, 
since it has an optimal decomposition in which all eight eigenstates of $\rho$ occur\cite{Siewert2012}. 
However, an extension to include up to three eigenstates in the decomposition would be enough to reproduce this
result.

For demonstrating how the algorithm works without manipulating the eigenbasis, I show the upper bound it offers
for the transverse Ising model. I show also a lower bound following Ref.~\cite{Eltschka2014},
which is positive only for $0\leq\lambda\lesssim0.9206$.
There are however strong hints that there is a non-vanishing threetangle 
in the transverse Ising model for nearest neighbors which will be close to the upper bound given here.
  
This might induce many possible directions of future research. 
One direction would be to think about further lower bounds to admit more than two parameters 
in the density matrices. Furthermore, one could try to enhance the number of 
eigenstates in the decomposition and improve the present algorithm.

\section*{Acknowledgements}

I acknowledge fruitful discussions with K. Krutitsky and R. Sch\"utzhold.


\end{document}